\documentclass[twocolumn,superscriptaddress]{revtex4}

\usepackage{amsmath}
\usepackage{graphicx}
\usepackage{amssymb}
\usepackage{multirow}

\DeclareRobustCommand{\stirling}{\genfrac\{\}{0pt}{}}
\DeclareMathOperator*{\argmax}{arg\,max}

\begin{document}

\title{Replicate immunosequencing as a robust probe of B cell repertoire diversity}

\author{William DeWitt}
\affiliation{Adaptive Biotechnologies Corporation, 1551 Eastlake Avenue East, Suite 200, Seattle, WA 98102}
\author{Paul Lindau}
\affiliation{University of Washington, Seattle, WA}
\affiliation{Fred Hutchinson Cancer Research Center, 1100 Fairview Ave N, Seattle, Washington 98109}
\author{Thomas Snyder}
\author{Marissa Vignali}
\author{Ryan Emerson}
\affiliation{Adaptive Biotechnologies Corporation, 1551 Eastlake Avenue East, Suite 200, Seattle, WA 98102}
\author{Harlan Robins}
\affiliation{Fred Hutchinson Cancer Research Center, 1100 Fairview Ave N, Seattle, Washington 98109}
\email[email:]{hrobins@fhcrc.org}

\begin{abstract}
Fundamental to quantitative characterization of the B cell receptor repertoire is clonal diversity - the number of distinct somatically recombined receptors present in the repertoire and their relative abundances, defining the search space available for immune response. This study synthesizes flow cytometry and immunosequencing to study memory and naive B cells from the peripheral blood of three adults. A combinatorial experimental design was employed, constituting a sample abundance probe robust to amplification stochasticity, a crucial quantitative advance over previous sequencing studies of diversity. These data are leveraged to interrogate repertoire diversity, motivating an extension of a canonical diversity model in ecology and corpus linguistics. Maximum likelihood diversity estimates are provided for memory and naive B cell repertoires. Both evince domination by rare clones and regimes of power law scaling in abundance. Memory clones have more disparate repertoire abundances than naive clones, and most naive clones undergo no proliferation prior to antigen recognition.
\end{abstract}

\maketitle

\section{Introduction}
Under threat from diverse pathogens having regeneration times orders of magnitude shorter than human lifetimes, the human adaptive immune system achieves specific response through random somatic recombination of sequence encoding specificity of antigen receptors on T and B lymphocytes. Recognition triggers an immune response, including cellular proliferation of the antigen-specific clone, recruitment to immunological memory, and, for B cells, production of antigen-specific antibodies. The repertoire of clones and their abundances evolve throughout an individual's life in response to exposures. Due to somatic hypermutation of the B cell receptor (BCR) triggered by antigen stimulation, B cells somatically evolve increasing antigen specificity as they proliferate. The repertoire of receptors is therefore extremely diverse.

Previous approaches to antigen receptor repertoire diversity estimation redeploy methods developed in the ecology and corpus linguistics literature to estimate species diversity and vocabulary size (see review \cite{Bunge:1993jh}), respectively. Specifically, Poisson abundance models, with both parametric and nonparametric estimators, are used. Although conceptually erroneous, mark-recapture formulae have also been applied \cite{Vollmers13082013}. Antigen receptor repertoires more closely achieve the idealizations of these models than the their original applications; populations are very large and well-mixed, and detection probabilities are homogeneous. However, studies suffer from limitations in sequencing data that blunt sophisticated computational approaches.

Robins et al. \cite{Robins:2009da} assessed T cell receptor (TCR) richness from high-throughput immunosequencing data using a nonparametric empirical Bayes method requiring divergent series regularization \cite{GOOD:1956fs, Efron:1976tt} (a substantially improved regularization technique, applied to estimating the molecular complexity of PCR libraries, is advanced in \cite{Daley:2013ef}). However, the sequencing read count assigned to each unique TCR (after error correction) was associated with its clonal abundance in the sample. This introduces noise and bias, since each single template is stochastically amplified by PCR. Although this high-throughput study captured the diversity of a realistic biological sample, inference of repertoire richness was problematic due to limited quantitation of sample abundance for each clone.

Rempala et al. \cite{Rempala:2011gm, Greene:2013jc} employed a likelihood model and posterior inference for mouse TCR richness using single-cell sequencing to quantitate sample abundance of T cell clones. Although this approach allows for precise quantitation of sample abundance, it is so low throughput (one cell per well on a 96-well plate) that diversity estimation was only possible for transgenic mice engineered to have dramatically limited TCR diversity. Although quantitatively principled, severe experimental limitations restricted the study to less biologically relevant repertoires.

\begin{figure*}
\begin{center}
\includegraphics[width=.9\linewidth]{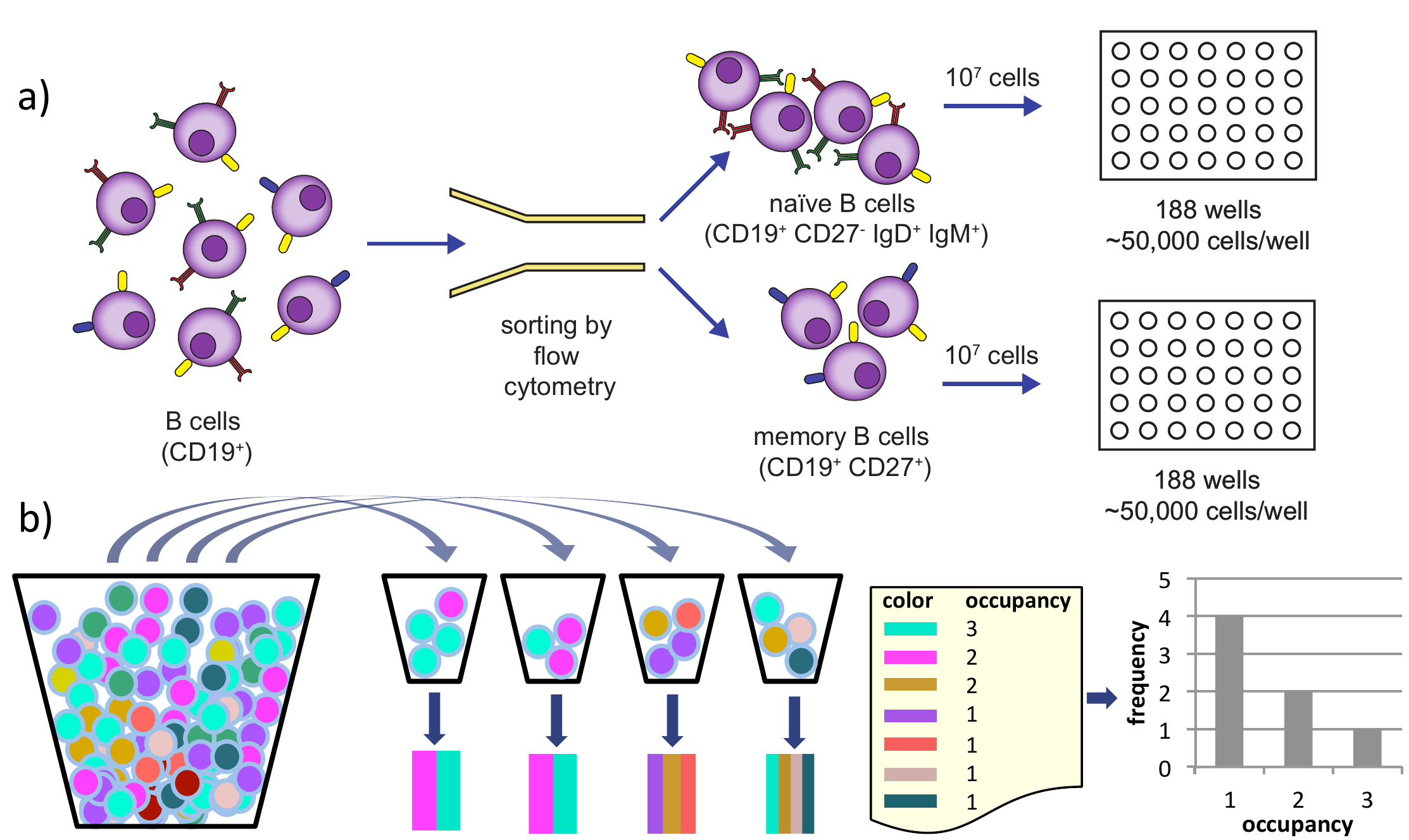}
\caption{\textbf{Experimental design. a)} Schematic of experimental design. B cells are sorted into memory and naive populations, and allocated evenly among PCR replicates. \textbf{b)} Schematic of urn model. Cells are colored balls with colors corresponding to clone identity. Each ball (cell) is randomly allocated to a sample bin (PCR replicate). Count censoring results in occupancy data.}
\label{schematic}
\end{center}
\end{figure*}

\section{Experimental Design}

In the present study a high-throughput and quantitatively robust (albeit indirect) probe of B cell clone sample abundance was devised. B cells from three adults were sorted into memory and naive populations (with two naive replicates for subject 1), each with $\sim10^7$ cells (Fig. \ref{schematic}a). Extracted DNA from each sample was evenly partitioned into 188 PCR replicates for amplification and uniquely barcoded for immunosequencing of the rearranged IgH locus \cite{Carlson:2013hd}, identifying clones by unique CDR3 sequence in their B cell receptor. Instead of relying on sequencing read counts to estimate a clone's sample abundance, we use its occupancy - the number of replicates it is observed in. In the regime of small occupancies, this approximates digital cell counting - a clone observed in only one replicate almost surely has a sample abundance of one cell. For larger sample abundances, co-occupancies become more probable, so occupancy increasingly underestimates abundance. Clones with sample abundance much larger than the number of replicates will saturate, appearing in all replicates.

To address possible template quantity variation across replicates and non-detection effects, we selected the subset of 150 replicates for each sample having minimum variance in the number of unique clones. Removing replicates with outlying allocations of cells or underperforming amplification is necessary to avoid breaking exchange symmetries invoked in our model.

\section{Model}

We advance a combinatorial extension of a well-studied model of sample abundance, enabling application to occupancy data. After introducing a parameterization of this extended model, a maximum likelihood diversity estimation is introduced, validated with simulations, and applied to BCR repertoire occupancy data to infer both richness (the number of clonal species) and an index of relative diversity (evenness of clone abundances).

\subsection{Poisson abundance model of replicate occupancy}

As is canon in the ecology and corpus linguistics literature, we begin by modeling sampling from a diverse population as a superposition of homogeneous Poisson processes. A mixing measure, $\mu(\lambda)$, characterizes the distribution of Poisson rates over all categories (B cell clones, as identified by productively rearranged IgH CDR3 segment, in our case). Since a clone's Poisson rate, $\lambda$, is given by its repertoire fraction times the number of cells sampled, $\mu(\lambda)$ is tantamount to the repertoire clonal abundance distribution. Homogeneity entails the approximation that the repertoire is effectively an infinite reservoir (or is being sampled with replacement), such that the data is not sensitive to depletion of the population fractions of the sampled clones. An equivalent urn model has an urn with an infinite number of balls, an unknown finite number of ball colors, and nonzero fractional abundances for each color. The total number of balls (cells) sampled from the urn (repertoire) is taken to be a Poisson sample from a multinomial population. The marginal distributions of sample cellular abundance, $j$, of each clone are then independently and identically distributed as
\[
p\left(j \middle\vert \mu(\lambda)\right) = \int_0^\infty d\mu(\lambda) \frac{\lambda^j e^{-\lambda}}{j!}
\]

To model replicate occupancy, we assume that each sampled cell is randomly assigned to one of $L$ possible replicates with equal probability (Fig. \ref{schematic}b). Because the sample material was partitioned equally among the $L$ replicates, it is not strictly correct to assume that each cell is assigned to a replicate independently. However, for large samples this approximation is very accurate. If the sample contains $N$ cells, then under this model the number of cells in each replicate is binomially distributed with $N$ trials and success probability $1/L$. The coefficient of variation is $\sqrt{(L-1)/N}$. The number of replicates used in this study was 150, and about 10 million cells were sequenced for all samples, leading to a coefficient of variation of about 0.004.

The distribution of a clone's replicate occupancy, $i$, conditioned on sample abundance, $j$, is then determined combinatorially as
\[
q\left(i \middle\vert j\right) = \frac{\binom{L}{i} i! \stirling{j}{i}}{L^j}.
\]
This is simply the ratio of the number of ways to partition $j$ cells into $i$ out of $L$ replicates, divided by the total number of ways to allocate $j$ cells among $L$ replicates. $\stirling{j}{i}$ denote Stirling numbers of the second kind, which count the number of ways to partition $j$ distinguishable objects into $i$ indistinguishable nonempty subsets.

Marginalizing over the hidden sample abundance gives the distribution of each clone's occupancy as
\begin{eqnarray}
r\left(i \middle\vert \mu(\lambda)\right) &=& \sum_{j=0}^\infty q\left(i|j\right) p\left(j \vert \mu(\lambda)\right) \label{model1} \\
&=& \sum_{j=0}^\infty \frac{\binom{L}{i} i! \stirling{j}{i}}{L^j} \int_0^\infty d\mu(\lambda) \lambda^j e^{-\lambda}/j! \nonumber\\
&=& \binom{L}{i} i! \int_0^\infty d\mu(\lambda) e^{-\lambda} \sum_{j=0}^\infty \frac{\stirling{j}{i}}{j!} \left(\frac{\lambda}{L}\right)^j \nonumber\\
&=& \binom{L}{i} \int_0^\infty d\mu(\lambda)e^{-\lambda} \left(e^{\frac{\lambda}{L}} - 1\right)^i, \nonumber
\end{eqnarray}
where we have exchanged the order of integration and summation, and identified the sum as a well-studied exponential generating function for the Stirling numbers (\cite{Wilf:478083}, p.83). In formal power series notation, $\stirling{j}{i} = j! \left[z^j\right]\left((e^z-1)^i/i!\right)$. 
We consider a finite-dimensional subspace of measures, $\mu_\theta(\lambda)$, parameterized by the vector $\theta$, and thus write \eqref{model1} as
\begin{eqnarray}
r_\theta(i) &=& \binom{L}{i} \int_0^\infty d \mu_\theta(\lambda) e^{-\lambda} \left(e^{\frac{\lambda}{L}} - 1\right)^i.
\label{model}
\end{eqnarray}

For a repertoire with clonal diversity $S$, the sample occupancy of each clone is drawn from distribution \eqref{model}. Let $l_1, l_2, \dots, l_S$ denote the replicate occupancies of $S$ labelled clones. For a very diverse repertoire and a limited sample, many clones will not be sampled, and thus have occupancy zero (the \textit{missing species}). Due to exchangeability of the clone labels, it is sufficient to consider the frequencies of nonzero occupancies, defined by the vector indicator random variable $o=(o_1, o_2,\dots, o_L)$, with $o_i = \left\vert\left\{c \in \left\{1,2,\dots, S\right\}: l_c=i\right\}\right\vert$ (the number of clones occupying exactly $i$ replicates).

We may write a multinomial likelihood function as
\begin{equation}
\mathcal{L}\left(\theta, S \middle\vert o\right) = \frac{S!}{(S-s)!} r_\theta(0)^{S-s}\prod_{i=1}^L \frac{r_\theta(i)^{o_i}}{o_i!}
\label{like}
\end{equation}
where $s = \sum_{i=1}^L o_i$ is the sample diversity. There are $S-s$ missing species.

\subsection{Parameterization}

Antigen receptor repertoires have been observed to follow Zipf's law \cite{Mora:2010wx}: the logarithms of the frequencies of clones are inversely proportional to the logarithms of their ranks by frequency. As a continuous analog of this discrete power law behavior, we make the parametric ansatz $d\mu(\lambda) \propto \lambda^{\gamma-1} \exp\left(-\frac{\lambda_a}{\lambda}-\frac{\lambda}{\lambda_b}\right)d\lambda$. The exponential factors cut off scaling behavior from below and above, and correspond to minimum and maximum abundances in the repertoire. This distribution, properly normalized, is the generalized inverse Gaussian \cite{Jorgensen:1982ct}. For Poisson abundance models, the parameters $\lambda_a$ and $\lambda_b$ are strongly asymptotically correlated in the likelihood for fixed $\gamma$ \cite{Stein:1987gy, Willmot:1988in}. This manifests as a ridge in parameter space that confounds likelihood maximization. A transformation that minimizes off-diagonal components of the Fisher information matrix is therefore introduced, resulting in the more orthogonal parameterization
\begin{equation}
d\mu_\theta(\lambda) = \frac{\xi^{-\gamma}}{2 K_\gamma(\omega)} \lambda^{\gamma-1} e^{-\frac{\omega}{2}\left(\frac{\xi}{\lambda} + \frac{\lambda}{\xi}\right)}d\lambda,
\label{GIG}
\end{equation}
with parameter vector $\theta = (\gamma, \omega, \xi)$. $K_\gamma(\omega)$ denotes the modified Bessel function of the second kind, arising by imposing normalization. Excellent fits to naive and memory occupancy data were obtained with mixtures of two such distributions (see section \ref{results}). Lognormal and Pareto distributions were also considered, but produced substantially worse results.

Under the parameterization \eqref{GIG}, the distribution \eqref{model} becomes
\begin{eqnarray}
r_\theta(i) &=& \binom{L}{i} \frac{\xi^{-\gamma}}{2 K_\gamma(\omega)} \int_0^\infty d\lambda \ \frac{\lambda^{\gamma-1} \left(e^{\frac{\lambda}{L}} - 1\right)^i}{e^{\lambda+\frac{\omega}{2}\left(\frac{\xi}{\lambda} + \frac{\lambda}{\xi}\right)}}.
\label{modelGIG}
\end{eqnarray}
Although not available in closed-form, these $L+1$ integrals can be approximated by quadrature to evaluate the likelihood \eqref{like}. Modeling as a mixture of two distributions of the form \eqref{GIG} adds a mixing parameter, $0\leq\alpha\leq1$, with $r_\theta(i) = (1-\alpha) r_{\theta_1}(i) + \alpha \ r_{\theta_2}(i)$.

\subsection{Maximum likelihood diversity estimation}

Direct maximization of the likelihood \eqref{like} is computationally formidable, as it constitutes a mixed integer nonlinear programming problem. However, it may be factorized in the suggestive form
\[
\mathcal{L}\left(\theta, S \middle\vert o\right) = \mathcal{L}_b\left(\theta, S \middle\vert o\right) \mathcal{L}_m\left(\theta \middle\vert o\right),
\]
where we define the binomial
\[
 \mathcal{L}_b\left(\theta, S \middle\vert o\right) = \binom{S}{s} r_\theta(0)^{S-s} \left(1-r_\theta(0)\right)^s,
 \label{binomial}
\]
and zero-truncated multinomial
\[
\mathcal{L}_m\left(\theta \middle\vert o\right) = s! \prod_{i=1}^L \frac{1}{o_i!}\left(\frac{r_\theta(i)}{1-r_\theta(0)} \right)^{o_i}.
\label{multi}
\]
An approach to approximate maximization of $\mathcal{L}\left(\theta, S \middle\vert o\right)$, proposed by Sanathanan as conditional maximum likelihood estimation \cite{Sanathanan:1977er}, is to first compute
\[
\hat{\theta} = \argmax_{\theta} \mathcal{L}_m\left(\theta \middle\vert o\right),
\]
which is independent of $S$ and can be obtained by nonlinear numerical maximization of the log-likelihood $\ell_m\left(\theta \middle\vert o\right)=\log \mathcal{L}_m\left(\theta \middle\vert o\right)$. A constrained gradient ascent algorithm \cite{Zhu:1997br} was used in the present work. Differentiation gives gradient components of the form
\[
\frac{\partial \ell_m\left(\theta \middle\vert o\right)}{\partial\theta_j} = \frac{s}{1-r_\theta(0)} \frac{\partial r_\theta(0)}{\partial\theta_j} + \sum_{i=1}^L \frac{o_i}{r_\theta(i)} \frac{\partial r_\theta(i)}{\partial\theta_j},
\]
with
\[
\frac{\partial r_\theta(i)}{\partial\theta_j} = \binom{L}{i} \int_0^\infty d \left(\frac{\partial\mu_\theta(\lambda)}{\partial\theta_j}\right) e^{-\lambda} \left(e^{\frac{\lambda}{L}} - 1\right)^i,
\]
which may be evaluated by quadrature for the parameterization \eqref{GIG}.

Having computed $\hat{\theta}$, it remains to maximize the richness piece of the likelihood. A lemma due to Chapman \cite{Chapman:QzHyJckJ} can be invoked to give
\begin{eqnarray*}
\hat{S} &=& \argmax_{S\in \mathbb{N}} \mathcal{L}\left(\hat{\theta}, S \middle\vert o\right)\\
&=& \argmax_{S\in \mathbb{N}} \mathcal{L}_b\left(\hat{\theta}, S \middle\vert o\right)\\
&=& \Bigl\lfloor\frac{s}{1-r_{\hat{\theta}}(0)}\Bigr\rfloor.
\end{eqnarray*}
Sanathanan's articulation of an asymptotic theory for the estimator $\hat{S}$ showed it to be equivalent to direct maximization of $\mathcal{L}\left(\theta, S \middle\vert o\right)$ for large $S$. A corresponding approach was taken by Rodrigues \cite{Rodrigues:2001kd} in an empirical Bayes treatment to approximate a posterior distribution for $S$. The density $\mu'_\theta(\lambda)$ is viewed as a prior which is realized in the repertoire for large $S$. Due to the large diversity of the BCR repertoire, we employ the diversity estimator $\hat{S}$, first investigating its accuracy via simulation.

\subsection{Shannon diversity in a Poisson abundance model}

To quantify the degree of uniformity in repertoire clonal abundance we derive a standard entropy-based index of diversity applied to a Poisson abundance model. For a repertoire with richness $S$ and clone-wise population fractions given by $\pi_1, \pi_2,\dots \pi_S$, the Shannon index \cite{Shannon:2001hi} is defined as the information entropy of the clone-wise abundance distribution.
\[
H = -\sum_{i=1}^S \pi_i \log\pi_i.
\]
The maximum entropy, $H_o = \log S$, occurs when $\pi_i = 1/S$ for all clones. The Shannon equitability is defined as $E_H = H/H_o = H/\log S$. $E_H$ ranges on the unit interval, with unity denoting maximally uniform abundance. It is a measure of disparity in abundance, with smaller values indicating more disparity.

In a Poisson abundance model, each clone, $i$, is assigned a Poisson frequency, $\lambda_i$, which is related to its population fraction, $\pi_i$, by $\lambda_i = \langle n \rangle \pi_i$, where $\langle n \rangle$ denotes the expected sample size.
\[
\langle n \rangle = \sum_{i=1}^S \lambda_i.
\]
With the measure parameterized by $\theta$ this becomes
\begin{eqnarray*}
\langle n \rangle_{S, \theta} &=& S \int_0^\infty d\mu_{\theta}(\lambda) \lambda \\
&=& S I_1(\theta),
\end{eqnarray*}
where we've defined the integral
\[
I_1(\theta) = \int_0^\infty d\mu_{\theta}(\lambda) \lambda.
\]

The Shannon index for a Poisson abundance model is then
\begin{eqnarray*}
H(S, \theta) &=& -S \int_0^\infty d\mu_{\theta}(\lambda) \frac{\lambda}{\langle n \rangle} \log \frac{\lambda}{\langle n \rangle} \\
&=& \log \langle n \rangle - \frac{S}{\langle n \rangle} \int_0^\infty d\mu_{\theta}(\lambda) \lambda \log \lambda \\
&=& \log S + \log I_1(\theta) - \frac{I_2(\theta)}{I_1(\theta)},
\end{eqnarray*}
with
\[
I_2(\theta) = \int_0^\infty d\mu_{\theta}(\lambda) \lambda \log\lambda.
\]
The equitability is then evaluated at the MLE as 
\begin{eqnarray*}
E_H(\hat{S}, \hat{\theta}) &=& \frac{H(\hat{S}, \hat{\theta})}{\log \hat{S}} \\
&=& 1 + \frac{1}{\log \hat{S}} \left(\log I_1(\hat{\theta}) - \frac{I_2(\hat{\theta})}{I_1(\hat{\theta})}\right)
\end{eqnarray*}
Finally, we define a measure of clonality as the complement of this quantity
\begin{eqnarray*}
C(\hat{S}, \hat{\theta}) &=& 1- E_H(\hat{S}, \hat{\theta})\\
&=& \frac{1}{\log \hat{S}} \left(\frac{I_2(\hat{\theta})}{I_1(\hat{\theta})}-\log I_1(\hat{\theta})\right).
\end{eqnarray*}
$C$ ranges from 0 (minimally clonal, equal abundance across clones) to 1 (maximally clonal).

For parameterization \eqref{GIG} the necessary integrals may be evaluated in terms of modified Bessel functions as
\[
I_1(\hat{\theta}) = \xi \frac{K_{-\gamma-1}(\omega)}{K_\gamma(\omega)}
\]
and
\[
I_2(\hat{\theta}) = \frac{\xi}{K_{\gamma }(\omega )} \left(\log \xi \ K_{\gamma +1}(\omega )-\left[ \frac{d}{dx} K_x(\omega )\right]_{x=-\gamma -1}\right)
\]
It is trivial to extended this to a model with two mixed generalized inverse Gaussians.

\section{Results}
\label{results}

\subsection{Simulation validation}

\begin{figure}
\begin{center}
\includegraphics[width=\linewidth]{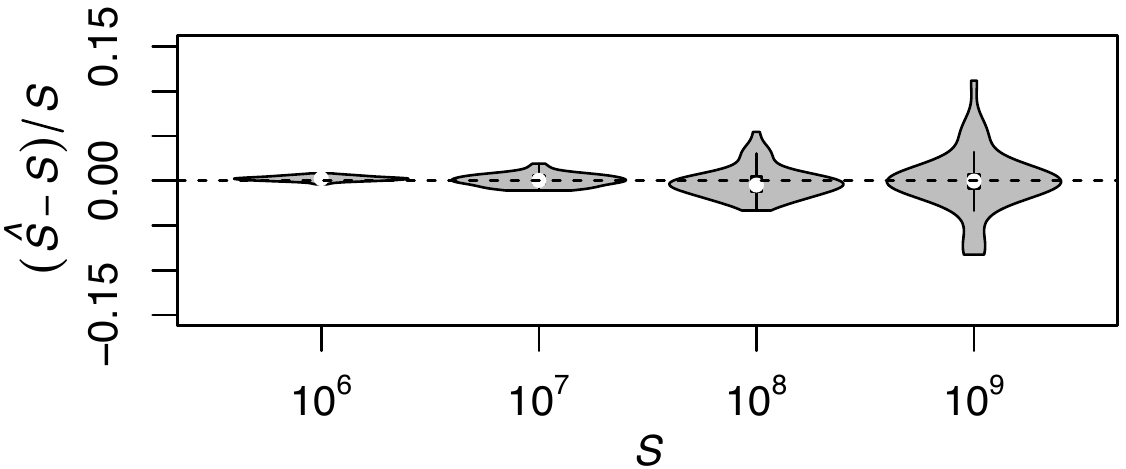}
\caption{\textbf{Performance of diversity estimation on simulated data.} One hundred simulations were performed for each of four diversity values, and Diversity estimates were computed for each. The resulting fractional errors are summarized as violin plots.}
\label{violin}
\end{center}
\end{figure}
To validate our methodology for inferring richness, simulations were performed by generating random draws from the likelihood \eqref{model}. Fig. \ref{violin} shows violin plots for fractional error in diversity estimation for four sets of 100 simulations. Each violin is for a set of 100 simulations with identical diversity ($S$) and shows the distribution of the fractional error of the MLE, $(\hat{S} - S)/S$.

The values of $S$ for the four sets are $10^6$, $10^7$, $10^8$, and $10^9$. For all four sets, the expected sample size is fixed at about 4.7 million cells. This is achieved by tuning the scale parameter $\xi$ inversely as $S$. This is necessary to address a property of the sampling model: the expected sample size is proportional to both $S$ and $\xi$, but we want expected sample size to be the same in all simulations as we tune $S$. Remaining parameters were fixed at $\gamma=-1$ and $\omega=0.01$ across all simulations (values similar to those arising in analyzing real data). Even at the high end of diversity, the expected error is only a few percent, demonstrating the efficacy of conditional maximum likelihood estimation in estimating an unknown population parameter.

\subsection{B cell diversity estimation}

\begin{figure*}
\begin{center}
\includegraphics[width=\linewidth]{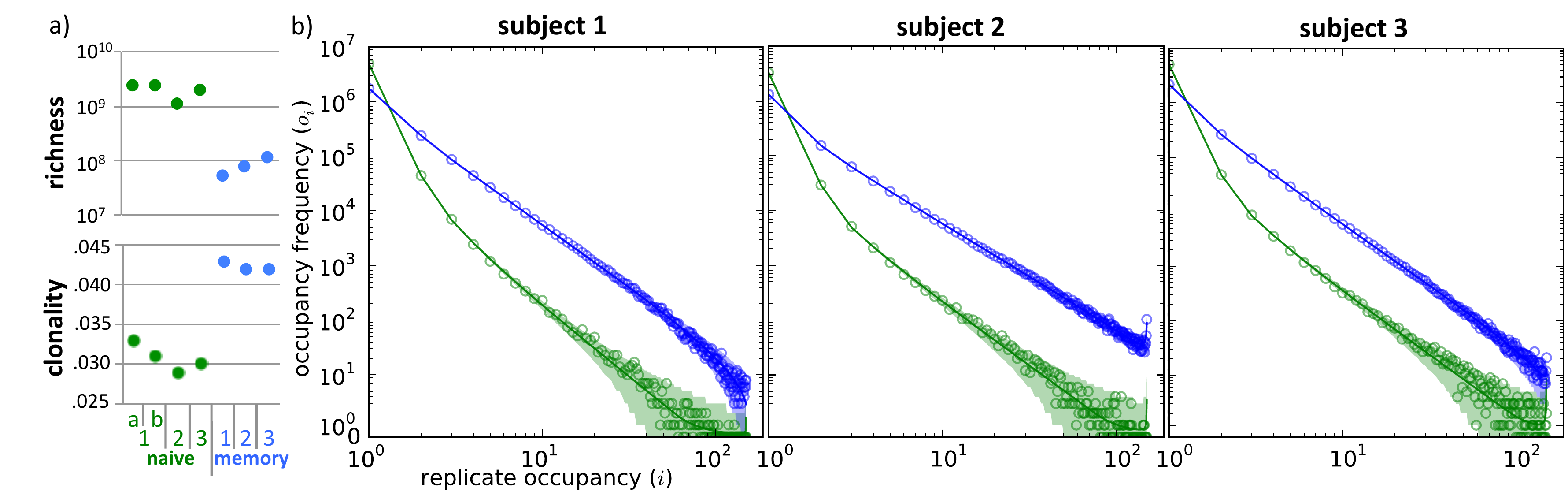}
\caption{\textbf{Diversity estimation results. a.)} Inferred richness and clonality indicates extreme naive diversity and clustering by cell type on both metrics. \textbf{b.)} Replicate occupancy data (circles) for naive (green) and memory (blue) samples, with expected occupancies at the MLE (solid lines) and 99\% marginal intervals (colored bands), indicating goodness of fit. Data for the second naive sample for patient 1 is omitted because results were not visually distinct. See Table \ref{MLEtable} for numerical details of MLE results.}
\label{data}
\end{center}
\end{figure*}
\begin{table*}
\caption{\textbf{MLE details.} Diversity, ($\hat{S}$), parameterization, $\hat{\theta} = \theta_1+\alpha\theta_2$, and clonality, $C(\hat{S}, \hat{\theta})$, for all samples.}
\resizebox{.95\linewidth}{!}{%
\begin{tabular}{|c|c|c|c|c|c|c|}
\hline
\begin{tabular}[c]{@{}c@{}}subject\end{tabular} & population       & $\hat{S}$ ($10^9$) & $\theta_1$             & $\theta_2$              & $\alpha$ & $C(\hat{S}, \hat{\theta})$ \\ \hline
\multirow{3}{*}{\textbf{1}}                         & \textbf{naive 1} & 2.39         & (-1.46, 9.04 10$^{-4}$, 8.69) & (-.097, .379 , 9.83 10$^{-4}$) & .944     & .033                         \\ \cline{2-7} 
                                                    & \textbf{naive 2} & 2.46         & (-1.47, 9.44 10$^{-4}$, 8.69) & (-.102, .410, 8.74 10$^{-4}$)  & .945     & .031                         \\ \cline{2-7} 
                                                    & \textbf{memory}  & .0527        & (-1.13, 7.98 10$^{-2}$, 9.07)   & (-.0722, .293, .0132)   & .960     & .043                         \\ \hline
\multirow{2}{*}{\textbf{2}}                         & \textbf{naive}   & 1.11         & (-1.23, 7.28 10$^{-3}$, 8.69) & (-.0944, .416, 1.68 10$^{-3}$) & .998     & .029                         \\ \cline{2-7} 
                                                    & \textbf{memory}  & .0765        & (-.860, 3.67 10$^{-2}$, 9.08) & (-.0706, .285, 6.38 10$^{-3}$) & .973     & .042                         \\ \hline
\multirow{2}{*}{\textbf{3}}                         & \textbf{naive}   & 1.97         & (-1.25, 1.80 10$^{-3}$, 8.69) & (-.0962, .393, 1.39 10$^{-3}$) & .989     & .030                         \\ \cline{2-7} 
                                                    & \textbf{memory}  & .116         & (-1.13, 5.03 10$^{-2}$, 9.07) & (-.0778, .287, 6.67 10$^{-3}$) & .968     & .042                         \\ \hline
\end{tabular}
}
\label{MLEtable}
\end{table*}

Results of the diversity estimation procedure described above are presented in Fig. \ref{data} and Table. \ref{MLEtable}. MLE richness and clonality inferences are shown in Fig. \ref{data}a. Occupancy data with visualized fits of the MLE are shown in Fig. \ref{data}b. Excellent fits are obtained for all data sets, as assessed by comparison to expectation values $\langle o_i \rangle = \hat{S}\  r_{\hat{\theta}}(i)$, $i=1,2,\dots, L$, and with variation characterized by the quantile functions of the binomial marginal of likelihood \eqref{like} at $\hat{S}$ and $\hat{\theta}$. Estimated richness, clonality, and parameterization values are very consistent between the two samples of subject 1's naive BCR repertoire, and take on characteristic values according to cell population. 

\section{Discussion}

By synthesizing flow cytometry and replicate immunosequencing, approximate digital cell counting of memory and naive B cell repertoires of three adults was enabled, providing the deepest and most quantitatively robust characterization of the repertoire yet available. Diversity of the repertoire was inferred using a novel likelihood model devised for replicate-based presence-absence data. Estimates of both clonal richness and evenness of abundance distributions were attained, showing consistency across individuals, but distinct clustering by cell population. Across naive samples, the estimated richness is similar to the expected number of total naive B cells in circulation, suggesting that the typical naive B cell undergoes no proliferation prior to antigen stimulation. Memory richness is consistent with several divisions on average, but higher disparity in abundance (indicated by lower clonality), likely corresponding to clonal expansions in response to antigen stimulation.

\begin{acknowledgments}
WD thanks Kameron Decker Harris for informative discussions and Erick Matsen for comments and corrections on an early draft of the manuscript.
\end{acknowledgments}

\bibliography{refs.bib}

\begin{thebibliography}{19}
\expandafter\ifx\csname natexlab\endcsname\relax\def\natexlab#1{#1}\fi
\expandafter\ifx\csname bibnamefont\endcsname\relax
  \def\bibnamefont#1{#1}\fi
\expandafter\ifx\csname bibfnamefont\endcsname\relax
  \def\bibfnamefont#1{#1}\fi
\expandafter\ifx\csname citenamefont\endcsname\relax
  \def\citenamefont#1{#1}\fi
\expandafter\ifx\csname url\endcsname\relax
  \def\url#1{\texttt{#1}}\fi
\expandafter\ifx\csname urlprefix\endcsname\relax\def\urlprefix{URL }\fi
\providecommand{\bibinfo}[2]{#2}
\providecommand{\eprint}[2][]{\url{#2}}

\bibitem[{\citenamefont{Bunge and Fitzpatrick}(1993)}]{Bunge:1993jh}
\bibinfo{author}{\bibfnamefont{J.}~\bibnamefont{Bunge}} \bibnamefont{and}
  \bibinfo{author}{\bibfnamefont{M.}~\bibnamefont{Fitzpatrick}},
  \bibinfo{journal}{J. Am. Statist. Assoc.} \textbf{\bibinfo{volume}{88}},
  \bibinfo{pages}{364} (\bibinfo{year}{1993}).

\bibitem[{\citenamefont{Vollmers et~al.}(2013)\citenamefont{Vollmers, Sit,
  Weinstein, Dekker, and Quake}}]{Vollmers13082013}
\bibinfo{author}{\bibfnamefont{C.}~\bibnamefont{Vollmers}},
  \bibinfo{author}{\bibfnamefont{R.~V.} \bibnamefont{Sit}},
  \bibinfo{author}{\bibfnamefont{J.~A.} \bibnamefont{Weinstein}},
  \bibinfo{author}{\bibfnamefont{C.~L.} \bibnamefont{Dekker}},
  \bibnamefont{and} \bibinfo{author}{\bibfnamefont{S.~R.} \bibnamefont{Quake}},
  \bibinfo{journal}{Proc. Natl. Acad. Sci. U.S.A.}
  \textbf{\bibinfo{volume}{110}}, \bibinfo{pages}{13463}
  (\bibinfo{year}{2013}).

\bibitem[{\citenamefont{Robins et~al.}(2009)\citenamefont{Robins, Campregher,
  Srivastava, Wacher, Turtle, Kahsai, Riddell, Warren, and
  Carlson}}]{Robins:2009da}
\bibinfo{author}{\bibfnamefont{H.~S.} \bibnamefont{Robins}},
  \bibinfo{author}{\bibfnamefont{P.~V.} \bibnamefont{Campregher}},
  \bibinfo{author}{\bibfnamefont{S.~K.} \bibnamefont{Srivastava}},
  \bibinfo{author}{\bibfnamefont{A.}~\bibnamefont{Wacher}},
  \bibinfo{author}{\bibfnamefont{C.~J.} \bibnamefont{Turtle}},
  \bibinfo{author}{\bibfnamefont{O.}~\bibnamefont{Kahsai}},
  \bibinfo{author}{\bibfnamefont{S.~R.} \bibnamefont{Riddell}},
  \bibinfo{author}{\bibfnamefont{E.~H.} \bibnamefont{Warren}},
  \bibnamefont{and} \bibinfo{author}{\bibfnamefont{C.~S.}
  \bibnamefont{Carlson}}, \bibinfo{journal}{Blood}
  \textbf{\bibinfo{volume}{114}}, \bibinfo{pages}{4099} (\bibinfo{year}{2009}).

\bibitem[{\citenamefont{Good and Toulmin}(1956)}]{GOOD:1956fs}
\bibinfo{author}{\bibfnamefont{I.~J.} \bibnamefont{Good}} \bibnamefont{and}
  \bibinfo{author}{\bibfnamefont{G.~H.} \bibnamefont{Toulmin}},
  \bibinfo{journal}{Biometrika} \textbf{\bibinfo{volume}{43}},
  \bibinfo{pages}{45} (\bibinfo{year}{1956}).

\bibitem[{\citenamefont{Efron and Thisted}(1976)}]{Efron:1976tt}
\bibinfo{author}{\bibfnamefont{B.}~\bibnamefont{Efron}} \bibnamefont{and}
  \bibinfo{author}{\bibfnamefont{R.}~\bibnamefont{Thisted}},
  \bibinfo{journal}{Biometrika} \textbf{\bibinfo{volume}{63}},
  \bibinfo{pages}{435} (\bibinfo{year}{1976}).

\bibitem[{\citenamefont{Daley and Smith}(2013)}]{Daley:2013ef}
\bibinfo{author}{\bibfnamefont{T.}~\bibnamefont{Daley}} \bibnamefont{and}
  \bibinfo{author}{\bibfnamefont{A.~D.} \bibnamefont{Smith}},
  \bibinfo{journal}{Nat. Meth.}  (\bibinfo{year}{2013}).

\bibitem[{\citenamefont{Rempala et~al.}(2011)\citenamefont{Rempala, Seweryn,
  and Ignatowicz}}]{Rempala:2011gm}
\bibinfo{author}{\bibfnamefont{G.~A.} \bibnamefont{Rempala}},
  \bibinfo{author}{\bibfnamefont{M.}~\bibnamefont{Seweryn}}, \bibnamefont{and}
  \bibinfo{author}{\bibfnamefont{L.}~\bibnamefont{Ignatowicz}},
  \bibinfo{journal}{J. Theor. Biol.} \textbf{\bibinfo{volume}{269}},
  \bibinfo{pages}{1} (\bibinfo{year}{2011}).

\bibitem[{\citenamefont{Greene et~al.}(2013)\citenamefont{Greene, Birtwistle,
  Ignatowicz, and Rempala}}]{Greene:2013jc}
\bibinfo{author}{\bibfnamefont{J.}~\bibnamefont{Greene}},
  \bibinfo{author}{\bibfnamefont{M.~R.} \bibnamefont{Birtwistle}},
  \bibinfo{author}{\bibfnamefont{L.}~\bibnamefont{Ignatowicz}},
  \bibnamefont{and} \bibinfo{author}{\bibfnamefont{G.~A.}
  \bibnamefont{Rempala}}, \bibinfo{journal}{J. Theor. Biol.}
  \textbf{\bibinfo{volume}{326}}, \bibinfo{pages}{1} (\bibinfo{year}{2013}).

\bibitem[{\citenamefont{Carlson et~al.}(2013)\citenamefont{Carlson, Emerson,
  Sherwood, Desmarais, Chung, Parsons, Steen, LaMadrid-Herrmannsfeldt,
  Williamson, Livingston et~al.}}]{Carlson:2013hd}
\bibinfo{author}{\bibfnamefont{C.~S.} \bibnamefont{Carlson}},
  \bibinfo{author}{\bibfnamefont{R.~O.} \bibnamefont{Emerson}},
  \bibinfo{author}{\bibfnamefont{A.~M.} \bibnamefont{Sherwood}},
  \bibinfo{author}{\bibfnamefont{C.}~\bibnamefont{Desmarais}},
  \bibinfo{author}{\bibfnamefont{M.-W.} \bibnamefont{Chung}},
  \bibinfo{author}{\bibfnamefont{J.~M.} \bibnamefont{Parsons}},
  \bibinfo{author}{\bibfnamefont{M.~S.} \bibnamefont{Steen}},
  \bibinfo{author}{\bibfnamefont{M.~A.} \bibnamefont{LaMadrid-Herrmannsfeldt}},
  \bibinfo{author}{\bibfnamefont{D.~W.} \bibnamefont{Williamson}},
  \bibinfo{author}{\bibfnamefont{R.~J.} \bibnamefont{Livingston}},
  \bibnamefont{et~al.}, \bibinfo{journal}{Nat. Comms.}
  \textbf{\bibinfo{volume}{4}} (\bibinfo{year}{2013}).

\bibitem[{\citenamefont{Wilf}(2000)}]{Wilf:478083}
\bibinfo{author}{\bibfnamefont{H.~S.} \bibnamefont{Wilf}},
  \emph{\bibinfo{title}{{Generatingfunctionology}}}
  (\bibinfo{publisher}{Academic Press}, \bibinfo{address}{London},
  \bibinfo{year}{2000}).

\bibitem[{\citenamefont{Mora et~al.}(2010)\citenamefont{Mora, Walczak, Bialek,
  and Callan}}]{Mora:2010wx}
\bibinfo{author}{\bibfnamefont{T.}~\bibnamefont{Mora}},
  \bibinfo{author}{\bibfnamefont{A.~M.} \bibnamefont{Walczak}},
  \bibinfo{author}{\bibfnamefont{W.}~\bibnamefont{Bialek}}, \bibnamefont{and}
  \bibinfo{author}{\bibfnamefont{C.~G.} \bibnamefont{Callan},
  \bibfnamefont{Jr}}, \bibinfo{journal}{Proc. Natl. Acad. Sci. U.S.A.}
  \textbf{\bibinfo{volume}{107}}, \bibinfo{pages}{5405} (\bibinfo{year}{2010}).

\bibitem[{\citenamefont{J{\o}rgensen}(1982)}]{Jorgensen:1982ct}
\bibinfo{author}{\bibfnamefont{B.}~\bibnamefont{J{\o}rgensen}},
  \emph{\bibinfo{title}{{Statistical Properties of the Generalized Inverse
  Gaussian Distribution}}}, vol.~\bibinfo{volume}{9} of
  \emph{\bibinfo{series}{Lecture Notes in Statistics}}
  (\bibinfo{publisher}{Springer}, \bibinfo{address}{New York, NY},
  \bibinfo{year}{1982}).

\bibitem[{\citenamefont{Stein et~al.}(1987)\citenamefont{Stein, Zucchini, and
  Juritz}}]{Stein:1987gy}
\bibinfo{author}{\bibfnamefont{G.~Z.} \bibnamefont{Stein}},
  \bibinfo{author}{\bibfnamefont{W.}~\bibnamefont{Zucchini}}, \bibnamefont{and}
  \bibinfo{author}{\bibfnamefont{J.~M.} \bibnamefont{Juritz}},
  \bibinfo{journal}{J. Am. Statist. Assoc.} \textbf{\bibinfo{volume}{82}},
  \bibinfo{pages}{938} (\bibinfo{year}{1987}).

\bibitem[{\citenamefont{Willmot}(1988)}]{Willmot:1988in}
\bibinfo{author}{\bibfnamefont{G.~E.} \bibnamefont{Willmot}},
  \bibinfo{journal}{J. Am. Statist. Assoc.} \textbf{\bibinfo{volume}{83}},
  \bibinfo{pages}{517} (\bibinfo{year}{1988}).

\bibitem[{\citenamefont{Sanathanan}(1977)}]{Sanathanan:1977er}
\bibinfo{author}{\bibfnamefont{L.}~\bibnamefont{Sanathanan}},
  \bibinfo{journal}{J. Am. Statist. Assoc.} \textbf{\bibinfo{volume}{72}},
  \bibinfo{pages}{669} (\bibinfo{year}{1977}).

\bibitem[{\citenamefont{Zhu et~al.}(1997)\citenamefont{Zhu, Byrd, Lu, and
  Nocedal}}]{Zhu:1997br}
\bibinfo{author}{\bibfnamefont{C.}~\bibnamefont{Zhu}},
  \bibinfo{author}{\bibfnamefont{R.~H.} \bibnamefont{Byrd}},
  \bibinfo{author}{\bibfnamefont{P.}~\bibnamefont{Lu}}, \bibnamefont{and}
  \bibinfo{author}{\bibfnamefont{J.}~\bibnamefont{Nocedal}},
  \bibinfo{journal}{ACM Trans. Math. Softw.} \textbf{\bibinfo{volume}{23}},
  \bibinfo{pages}{550} (\bibinfo{year}{1997}).

\bibitem[{\citenamefont{Chapman}(1951)}]{Chapman:QzHyJckJ}
\bibinfo{author}{\bibfnamefont{D.}~\bibnamefont{Chapman}},
  \emph{\bibinfo{title}{{Some properties of the hypergeometric distribution
  with applications to zoological sample censuses}}}, University of California
  publications in statistics (\bibinfo{publisher}{University of California
  Press}, \bibinfo{year}{1951}).

\bibitem[{\citenamefont{Rodrigues et~al.}(2001)\citenamefont{Rodrigues, Milan,
  and Leite}}]{Rodrigues:2001kd}
\bibinfo{author}{\bibfnamefont{J.}~\bibnamefont{Rodrigues}},
  \bibinfo{author}{\bibfnamefont{L.~A.} \bibnamefont{Milan}}, \bibnamefont{and}
  \bibinfo{author}{\bibfnamefont{J.~G.} \bibnamefont{Leite}},
  \bibinfo{journal}{Biom. J.} \textbf{\bibinfo{volume}{43}},
  \bibinfo{pages}{737} (\bibinfo{year}{2001}).

\bibitem[{\citenamefont{Shannon}(1948)}]{Shannon:2001hi}
\bibinfo{author}{\bibfnamefont{C.}~\bibnamefont{Shannon}},
  \bibinfo{journal}{Bell Syst. Tech. J.} \textbf{\bibinfo{volume}{27}},
  \bibinfo{pages}{379} (\bibinfo{year}{1948}).

\end{thebibliography}

\end{document}